\pdfoutput=1

\documentclass[11pt]{article}

\usepackage[utf8]{inputenc}
\usepackage[T1]{fontenc}
\usepackage[a4paper,margin=2.5cm]{geometry}
\usepackage{graphicx}
\usepackage{amsmath}
\usepackage{url}

\usepackage{lineno}
\usepackage{xcolor}

\usepackage[superscript]{cite} 
\title{A magnetic environment with reproducible spatio-temporal magnetic conditions at picotesla level}

\author{
Philipp Wunderl$^{1,*,\dagger}$, Maximilian Huber$^{1,\dagger}$, Florian Büttiker$^{4}$, Jonas Emrich$^{2}$, \protect\\
Peter Ewert$^{3}$, Peter Fierlinger$^{1}$, Reinhard Heckel$^{2}$, Felix Herz$^{1}$, Tobias Jensch$^{1}$, \protect\\
Kim Kölbl$^{1}$, Florian Kuchler$^{1}$, Thorben Lützel$^{1}$, Karin Narushima$^{3}$, René Seiger$^{1}$, \protect\\
Jonas Vögtle$^{1}$, Annette Wacker-Gussmann$^{3}$, Chiara Weckmann$^{1}$, Lena Wunderl$^{1}$, \protect\\
Peisen Zhao$^{1}$
}

\begin{document}
\maketitle

\vspace{-1.5em}
\noindent{\footnotesize{
$^{1}$Technical University of Munich, TUM School of Natural Sciences, Paula-Hahn-Weinheimer-Straße 1, 85748 Garching, Germany\protect\\
$^{2}$Technical University of Munich, TUM School of Computation, Information and Technology, Arcisstraße 21, 80333 Munich, Germany\protect\\
$^{3}$Clinic of Congenital Heart Disease and Pediatric Cardiology, German Heart Center, TUM University Hospital, Technical University of Munich, Lazarettstraße 36, 80636 Munich, Germany\protect\\
$^{4}$IMEDCO AG, Industriestrasse West 14, 4614 Hägendorf, Switzerland\protect\\
\\
$^{\dagger}$These authors contributed equally to this work.\\
$^{*}$ e-mail: philipp.wunderl@tum.de
}}

\begin{abstract}
Walk-in, picotesla-scale environments are essential for measurements of biomagnetism and fundamental physics. However,  conventional rooms require typical active dynamic compensation, which adds complexity and magnetic noise. Here, we present a solution that achieves an absolute residual field well below 100$\,$pT within its central measurement volume. Following magnetic equilibration, this environment achieves picotesla-scale reproducibility. Consequently, optically pumped magnetometers operate at their design noise and drift performance and remain operational during sensor motion without active feedback. A key technique is robotic mapping, which resolves the ultra-low residual field patterns and demagnetization stability. We demonstrate the platform's versatility through high-fidelity adult and fetal magnetocardiography, standing magnetoencephalography, and ultra-low field magnetic resonance with polarized noble gases in the limit of strongly coupled spins in a negligible holding field. The achieved passive reproducibility turns the ultra-low magnetic background into a predictable, correctable property, establishing a novel foundation for next-generation quantum sensing and precision physics.

\end{abstract}

\section*{Introduction}

Magnetically shielded rooms (MSRs) constitute the essential infrastructure for high-precision biomagnetic measurements and fundamental physics experiments. Since the pioneering developments by Cohen in the late 1960s \cite{cohen1970large,cohen1970mcg}, the architecture of multi-layer high-permeability shielding has undergone continuous refinement. Key milestones include the original Berlin MSR \cite{mager1981bmsr} and the subsequent eight-layer BMSR-2 at the Physikalisch-Technische Bundesanstalt (PTB) \cite{bork2000bmsr2}, which established shielding factors exceeding $10^7$ and residual magnetic flux densities below 500$\,$pT within a walk-in volume. Historically, MSR performance has been summarized by frequency-dependent shielding factors, achieved through flux-guiding in high-permeability layers and eddy-current damping in conductive shells \cite{bork2000bmsr2,thiel2007demagnetization,kadenShielding}.

However, for many modern applications, the shielding factor alone is an only partial suited and also incomplete metric \cite{ayres2022n2edm,ayres2024ultralow,altarev2014msr,altarev2015minimizing,thiel2007demagnetization}. Biomagnetic imaging, low-field nuclear spin precession and electric-dipole-moment (EDM) searches require not only the attenuation of external disturbances but also a minimized residual magnetic field, small spatial gradients, low intrinsic noise and, critically, a highly reproducible magnetic state.

In the realm of biomagnetism, the emergence of optically pumped magnetometers (OPMs) has established cryogen-free magnetoencephalography (MEG) and fetal magnetocardiography (fMCG) \cite{boto2018moving,brookes2022opm,budker2007optical}. While OPMs offer high sensitivity, they operate within a limited dynamic low field range and are susceptible to residual gradients during subject motion \cite{boto2018moving,holmes2022lightweight,holmes2023ambulatory,brookes2022opm}. A passive background field that is both small and which can be achieved in a reproducable way allows researchers to distinguish between sensor offsets, biological source signals and environmental drifts without the complexity of continuous active feedback, which is also always limited by the sensor performance. In cardiac measurements, such reproducibility enables quantitative comparisons between repeated maps, while in fetal measurements it provides the stable, low-noise environment necessary for the development of source-geometry models across different gestational ages \cite{wacker2022contribution,wacker2022jacc,wurm2023smallscale}.

Precision spin-precession experiments impose a complementary set of stringent requirements \cite{cates1988relaxation,chupp2019edm}. In EDM searches and Lorentz-invariance tests via spin-precession co-magnetometry, magnetic gradients directly contribute to transverse relaxation and systematic frequency shifts. Furthermore, particles moving in an electric field sample a motional magnetic field.
In the presence of magnetic-field gradients, this interaction can generate geometric-phase-induced frequency shifts that mimic an EDM signal \cite{chupp2019edm,pendlebury2004geometric,harris2006dipole}. An MSR whose residual field pattern can be precisely mapped, reproduced and corrected allows for a fundamentally different operating mode: the magnetic state is treated as a calibrated, deterministic object rather than an arbitrary initial condition.
Despite advancements in large-scale facilities, which reached residual-field levels below 300$\,$pT \cite{ayres2022n2edm,ayres2024ultralow}, systems achieving the lowest absolute fields have historically relied on active compensation like complex auxiliary coil systems. Comparable picotesla-scale cycle-to-cycle reproducibility of the full residual-field pattern in a passive walk-in environment has not been demonstrated \cite{ayres2022n2edm,ayres2024ultralow,altarev2014msr,altarev2015minimizing,thiel2007demagnetization}.

Here, we present a large-volume passive MSR designed to bridge the requirements of both biomagnetism and precision spin physics. The facility provides a walk-in experimental volume equipped with non-magnetic infrastructure, specialized optical lighting and an automated versatile magnetic equilibration system. Using a non-magnetic field measurement apparatus, we characterize the spatial field and temporal stability of the room. We demonstrate the scientific implications of this high reproducibility through several benchmarks: simultaneous adult magnetocardiography (MCG) and electrocardiography (ECG), raw fMCG signals at 21 weeks of gestation, finger-tapping MEG while the subject stands freely, without active field compensation and $^3$He precession at ultra-low holding fields of a few hundred picotesla. 
Finally, we discuss how this deterministic magnetic state enables the acquisition of high-fidelity, artifacts-free biomagnetic reference data by eliminating the need for active-feedback compensation, while the  reproducible state enables static corrections to suppress Berry-phase-induced systematics in EDM searches and facilitates the separation of sensor drift from true magnetic-field changes. In addition, we discuss how such a facility can be used to conduct research in exotic physics, such as the search for dark matter in new parameter spaces. This multifunctional utility underscores the room’s role as a calibrated metrological platform that not only provides an ideal environment for validating biophysical models but also enables a new generation of experiments in fundamental physics.

\section*{Results}

\subsection*{Achieving a highly reproducible passive magnetic field environment in the picotesla range}
The facility was designed to combine ultra-low residual fields with the practical requirements of human biomagnetic experiments. The shield is a rectangular walk-in enclosure with external dimensions of 3550$\,$mm $\times$ 3550$\,$mm $\times$ 3400$\,$mm and an inner mu-metall shield dimension of 2200$\,$mm $\times$ 2200$\,$mm $\times$ 2100$\,$mm. Four high-permeability Magnifer layers with nominal thicknesses of 2, 4, 4 and 3$\,$mm provide the low-frequency shielding, while an aluminium shell provides eddy-current and radio-frequency attenuation. The floor is walkable and can withstand heavy loads, the ceiling lighting is fibre coupled, and air, cable and gas feedthroughs are implemented as waveguides to preserve radio-frequency shielding.

The door is a critical element of the design because door overlaps can dominate the remanent field close to the opening. The two inner magnetic layers are integrated into a sliding door, while the two outer layers are operated as a swing door. This configuration gives a clear opening of approximately 1000$\,$mm $\times$ 2000$\,$mm and is engineered to shield the innermost layer from the Earth's magnetic field during ordinary operation. A lifting platform in front of the door allows equipment and participants to enter at-grade.

Magnetic equilibration is performed with coil sets mounted around the innermost and second-innermost shielding layers, complemented by internal coils generating homogenuous fields for experiments. For the measurement in this work we decided to use only the equilibration coils around the edges. \cite{altarev2014msr,altarev2015minimizing,allmendinger2023degaussing, thiel2007demagnetization} The modeling and simulation of shield demagnetization have already been thoroughly investigated in a previous study, and the findings have been applied here. \cite{sun2021limits}

A self-constructed automated switching unit cycles through coil configurations with direct-current relays, monitors current and voltage, and returns the drive current safely to zero without jumps. The best residual-field performance is obtained with a sequential equilibration procedure applied first to the innermost layer, then to the second-innermost layer and finally again to the innermost layer.

\begin{figure}[htbp]
\centering
 \includegraphics[width=1\linewidth]{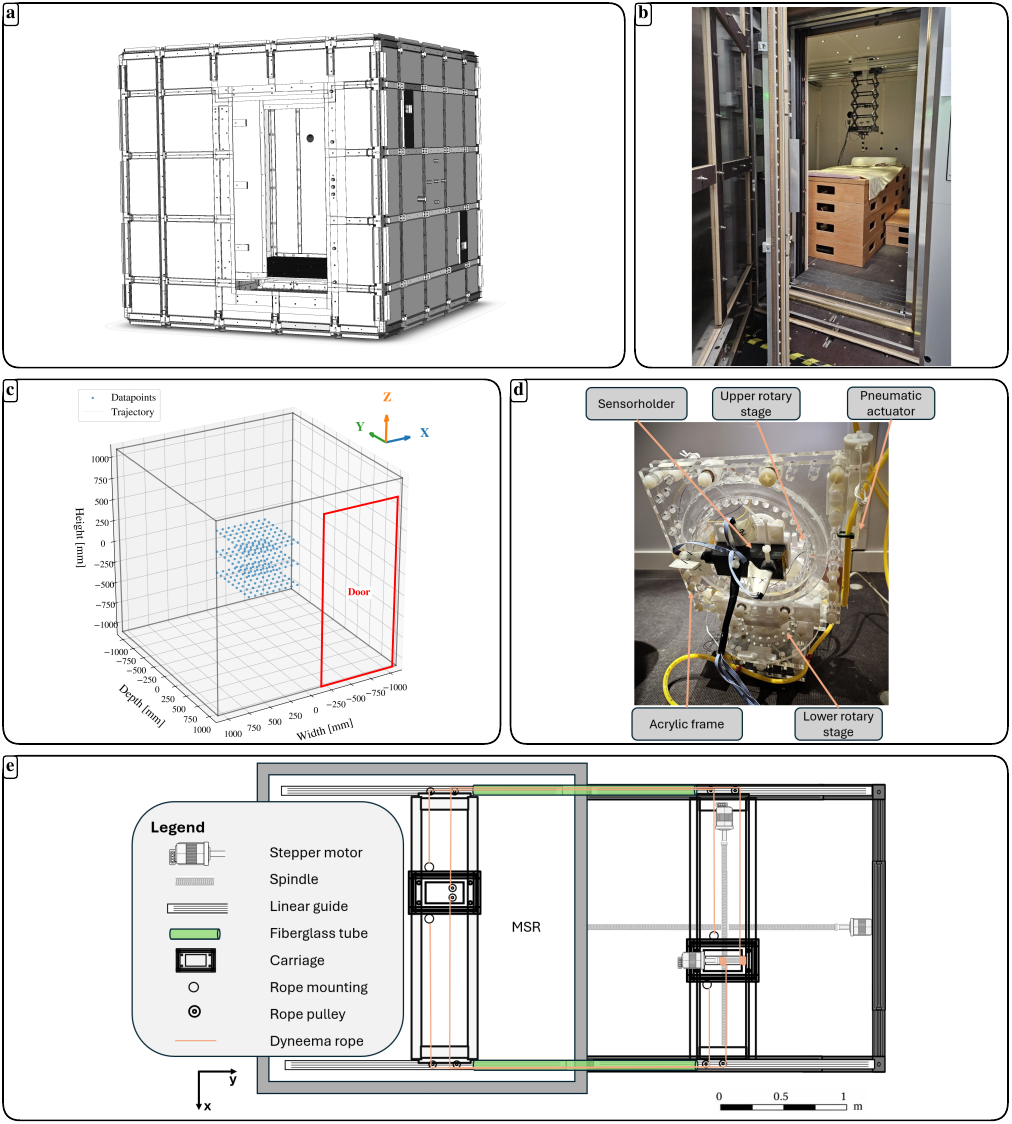}\
\caption{\textbf{Magnetically shielded room and non-magnetic characterization and calibration systems.} \textbf{a}, Simplified technical drawing of the MSR. The drawing shows the room with the sliding door open, without the swing door.  \textbf{b}, Image of the open door of the MSR with a bed set up for MCG measurements. \textbf{c} , Diagram of the measurement points for a typical measurement using the magnetic field mapping robot in the MSR.  \textbf{d}, Picture of the automated non-magnetic sensor-rotation mount used for offset calibration. The biaxial system is constructed from polymethyl methacrylate and uses pneumatic actuation with elastic drive bands to rotate the sensor about two orthogonal axes. \textbf{e}, Schematic top view of the robotic field-mapping device. The internal carriage is driven by Dyneema\textregistered~ropes (orange) via rope pulleys and fiberglass tubes (green). All active electronic components, including stepper motors and lead-screw spindles, are situated on an external support frame outside the MSR.}
\label{fig:setup}
\end{figure}

To characterize the resulting field with picotesla sensitivity, we developed a robotic mapping system in which all motors and control electronics remain outside the shielded volume (Fig.~\ref{fig:setup}). Mechanical motion is transferred into the room via fiberglass rods and Dyneema\textregistered~ropes, while the internal carriage accommodates either a triaxial fluxgate sensor or triaxial zero-field OPMs. Dyneema\textregistered~ropes are used because of their low elasticity to minimize backlash as much as possible. To ensure rigorous sensor-offset calibration, we integrated an automated non-magnetic rotation mount. The mount uses turntables actuated by elastic bands and compressed air to enable precise rotations about two orthogonal axes. This geometry and calibration methodology are essential for the performance claims presented here: the mapping system resolves the magnetic state of the room with no ferromagnetic or electrically active components inside the measurement volume, which also allows the mapper to remain permanently installed for all measurements.

\subsection*{Shielding factor characterization}

The frequency-dependent shielding factor was characterized by applying a known external alternating field through a large-scale coil system surrounding the MSR, with the internal response monitored by a stationary OPM sensor. Measurements across $10^{-3}$--$10^2$$\,$Hz reveal two distinct shielding regimes \cite{bork2000bmsr2,thiel2007demagnetization,kadenShielding}. In the quasi-static and low-frequency limit below 0.1$\,$Hz, the MSR maintains a shielding factor exceeding $6\times 10^4$. As the frequency increases toward 1$\,$Hz, the shielding factor rises sharply, peaking at values above $10^7$. This enhancement is attributed to the combined action of high-permeability shielding and eddy-current damping in the aluminium layers.

This attenuation is preserved even at the lowest measured excitation amplitudes, using a 0.25$\,$A coil current corresponding to $\mu$T-level external fields. This consistency confirms that the shielding performance in the small-signal regime, which is most relevant for suppressing environmental noise, is not compromised by nonlinear permeability of the shielding material. A complementary analysis at 0.01$\,$Hz as a function of excitation amplitude shows a nonlinear improvement in performance: the shielding factor rises from approximately 60,000 at a peak-to-peak amplitude of 100$\,$nT to more than 200,000 at large amplitudes near 100$\,$$\mu$T. This behaviour is consistent with the nonlinear permeability of high-permeability alloys in the low-field Rayleigh regime of the hysteresis loop, where effective permeability is enhanced \cite{fiorillo2004magnetic}. Consequently, the MSR exhibits an amplitude-dependent shielding response for large magnetic disturbances, which is advantageous for operation in magnetically noisy urban laboratory environments.

\subsection*{Residual field and field near door overlaps}
\begin{figure}[htbp]
\centering
 \includegraphics[width=1\linewidth]{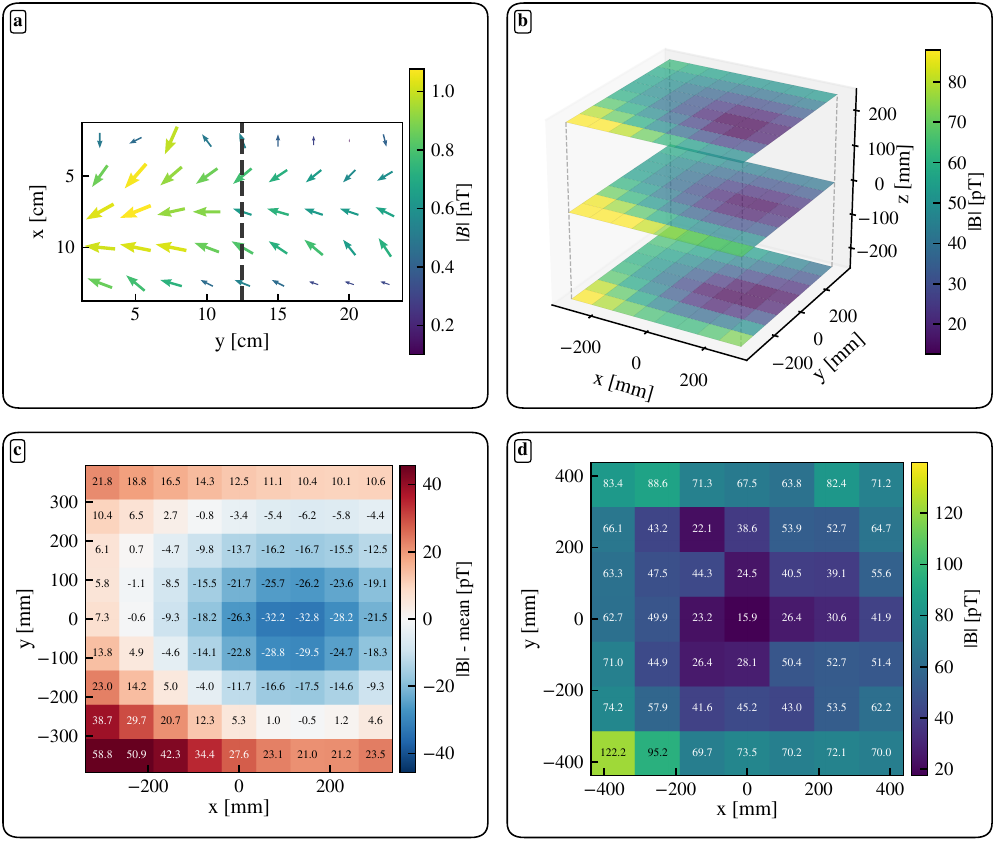}\
\caption{\textbf{Spatial characterization of the residual magnetic field.} \textbf{a}, Magnetic-field vector map adjacent to the door-overlap region. The door is located at y = 0, and the position of the overlap with the wall is indicated by the dashed line. Arrow direction indicates field orientation, while colour denotes magnitude in nT. The distortion is localized and approaches the central-room background within $\sim 6$$\,$cm of the seam.
\textbf{b}, Three-dimensional heatmap of the residual field magnitude $|B|$ across five horizontal planes following a single equilibration run, measured with an OPM. The field remains below 90$\,$pT throughout the measurement volume.
\textbf{c}, The homogeneity of the magnetic field at the mean altitude is calculated from the difference between the magnetic field at the measurement point and the mean value for the plane, measured with an OPM. The field is homogeneous here within 80$\,$pT. 
\textbf{d}, Comparison map generated using a three-axis Fluxgate sensor at the mid-level. The values systematically match those of the Quspin map within the uncertainty range.}
\label{fig:residual}
\end{figure}

Following magnetic equilibration (see Methods), the residual magnetic flux density $|B|$ was mapped  automated throughout the central volume and near the door interface. Figure~\ref{fig:residual}b presents stacked heatmaps of $|B|$ across multiple horizontal planes. Throughout the central measurement volume, the field magnitude remains below 90$\,$pT. The observed spatial structure reflects the geometry of the edge-only equilibration coils: the lowest values, 15--20$\,$pT, are concentrated in the lower-central region, while values of 80--90$\,$pT occur at the corners where the degaussing coils are least effective. This deterministic spatial pattern provides a stable baseline for potential static-field corrections.

The primary source of local field elevation in an MSR with an opening panel is the door seam, as panel overlaps are magnetized differently from the bulk regions and are more difficult to equilibrate perfectly \cite{altarev2015minimizing}. Figure~\ref{fig:residual}a shows a vector map in the horizontal plane immediately adjacent to the door overlap. While the field reaches approximately 1$\,$nT directly at the seam, the distortion is localized, falling below 400$\,$pT within 3$\,$cm and reaching the background level, below 200$\,$pT, within 6$\,$cm. For the subsequent measurements reported here, sensors were positioned at least 20$\,$cm from the door plane, making the door contribution negligible relative to the stable background.

\subsection*{Equilibration reproducibility}

\begin{figure}[htbp]
\centering
\centering
 \includegraphics[width=1\linewidth]{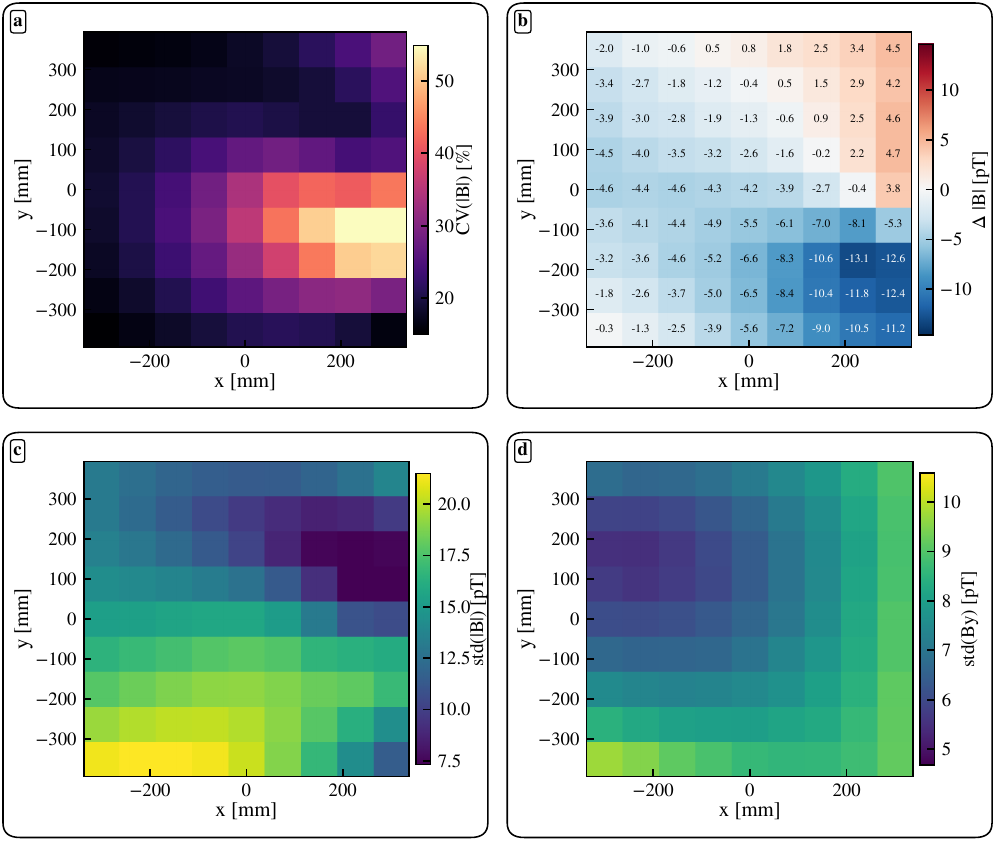}\
\caption{\textbf{
Quantitative reproducibility of the equilibrated magnetic state.} 
\textbf{a}, Statistical analysis over five independent equilibration cycles, showing the coefficient of variation $CV(|B|)$ and the standard deviation $\sigma(|B|)$. Map-level standard deviations of 8--20$\,$pT, together with an offset reproducibility approaching $\sim 5$$\,$pT, demonstrate picotesla-scale reproducibility of the residual-field pattern.
\textbf{b}, Point-by-point difference maps of residual-field components $d|B|$,  between two equilibration runs acquired on different days with intervening door operations. The systematic structure in the maps confirms that the residual field is a deterministic and reproducible property of the equilibration procedure.
\textbf{c}, Standard deviation of the absolute magnetic field at each point across the five maps. 
\textbf{d}, As an example the standard deviation of the magnetic field in y-direction at each point across the five maps. 
}
\label{fig:repro}
\end{figure}

The defining achievement of this facility is not merely the attainment of a low absolute field, but the reproducibility of the residual magnetic-field pattern across independent equilibration cycles. To assess this, we acquired point-by-point difference maps between successive equilibration runs, including consecutive directly runs and runs separated by door-opening events and multiple days. As shown in Fig.~\ref{fig:repro}a, differences between two consecutive equilibrations in individual field components are restricted to the picotesla scale. We observed variations from 0.0 to 4.5$\,$pT in $B_z$, from 3.1 to 10$\,$pT in $B_x$ and from -20 to 6.5$\,$pT in $B_y$ across the mapped area. These differences do not manifest as random point-to-point noise but as smooth component-wise patterns, suggesting that the dominant variations arise from subtle, stable shifts in the equilibration state or the absolute position of the mapper rather than stochastic fluctuations.

To quantify this behaviour over repeated trials, we computed the standard deviation $\sigma$ and the coefficient of variation, $CV=\sigma/\mu$, of the field magnitude $|B|$ over five independent equilibrations (Fig.~\ref{fig:repro}b) with $\mu$ the mean of the field magnitude. The standard deviation remains within the 8-20$\,$pT range throughout the mapped volume. Even where the absolute field is smallest, leading to a denominator effect in the coefficient of variation, the coefficient remains below 50\%. The lowest values are likely limited by sensor-offset reproducibility, confirmed at approximately 5$\,$pT, and long-term instrumentation drift.

This stability allows the passive achieved residual field to be treated as a deterministic and correctable background rather than an unpredictable initial condition. In practice, a static field correction derived from the mean of several equilibration runs can reduce the effective residual gradient to the level of a few pT/m over 10$\,$cm. This improves the coherence budget for spin-precession experiments and provides a reliable environment for repeated biomagnetic mappings under nominally identical magnetic conditions.

\subsection*{Room and sensor noise floor, long-term stability and motion tolerance}

A critical validation of any ultra-low field environment is whether magnetic sensors can operate at their fundamental performance limits without degradation from residual-field fluctuations, technical interference or steep gradients. We evaluated several OPM types within the MSR, including QuSpin QZFM Gen-2 and Gen-3 sensors and a Twinleaf MicroSERF sensor \cite{quspinQZFM,twinleafMicroSERF}. In all cases, the sensors operated at their manufacturer-specified noise floors across the relevant frequency range, without active shielding or parameter adjustment.

\begin{figure}[htbp]
\centering
 \includegraphics[width=1\linewidth]{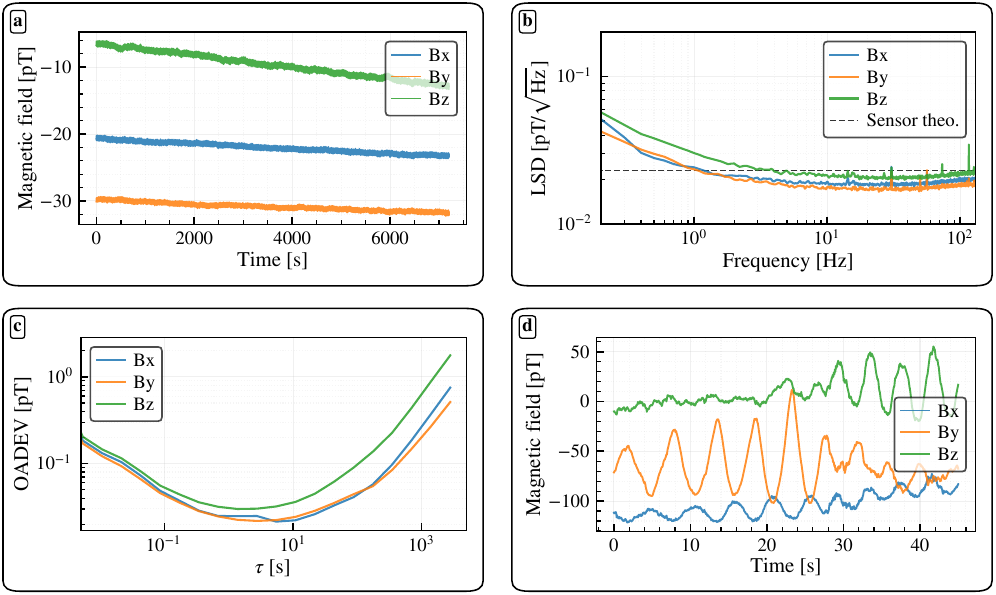}\
\caption{\textbf{Sensor noise, long-term stability and motion tolerance.} \textbf{a}, Two-hour OPM measurement used for drift and noise analysis using a QZFM sensor. For visualization, the data were downsampled to 50$\,$Hz. \textbf{b}, Linear spectral density of the three magnetic-field components recorded with a stationary OPM at the room centre, reaching the design noise floor above 3$\,$Hz. \textbf{c}, Overlapping Allan deviation identifying the optimum averaging time of $\sim 1$--5$\,$s and characterizing long-term drift. Based on this measurement alone, it is difficult to distinguish the sensor's drift from that of the environment, which is why it is only possible to speak of the combined drift of room and intrinsic drift of sensor at this point. However, subsequent helium precession measurements reveal that the drift observed here appears to be caused primarily by the sensors. \textbf{d}, Field components recorded during manual sensor motion at $\sim 20$$\,$cm/s. The signals remain within about $\pm 100$$\,$pT, demonstrating that the passive gradient is sufficiently low to preserve the sensor dynamic range during movement.}
\label{fig:noise}
\end{figure}

Representative linear spectral densities for a triaxial QuSpin QZFM are shown in Fig.~\ref{fig:noise}b, which were calculated from a two-hour measurement (see Fig.~\ref{fig:noise}a). All three field components approach the specified noise floor of $\sim 15$--23$\,$fT$\sqrt{\mathrm{Hz}}^{-1}$ above 3$\,$Hz. The absence of narrowband interference lines confirms that the shielded environment is free of technical noise from building infrastructure within the primary neuroscience frequency band, 1-100$\,$Hz. This performance across multiple OPM sensors demonstrates that the MSR provides a suitable zero-field environment for current-generation quantum sensors.

To quantify long-term stability, we computed the overlapping Allan deviation for all three components (Fig.~\ref{fig:noise}c). The stability curves reach a minimum of $\sim 30$--50$\,$fT at averaging times of $\tau \approx 1$--5$\,$s, defining the practical floor for quasi-static measurements. The gradual rise at longer intervals, $\tau>10$$\,$s, reflects slow relaxation of the magnetic state after equilibration as well as potential intrinsic sensor drift. For spin-precession experiments, these curves define the operational window and the timescales on which background drifts should be monitored or corrected.

Finally, we demonstrated the practical motion tolerance of the passive environment. Movement through a residual gradient converts spatial inhomogeneity into a time-dependent field at the sensor. Figure~\ref{fig:noise}d shows the field recorded while a sensor was moved by hand at approximately 20$\,$cm/s through the room interior. All three components remained within about $\pm 100$$\,$pT throughout the motion, well within the $\pm 5$$\,$nT dynamic range of the zero-field OPM. This confirms that the residual gradients are sufficiently small to allow typical participant motion without requiring active compensation to maintain linear sensor operation.

\subsection*{Simultaneous MCG mapping and ECG as a controlled cardiac reference measurement}

As a primary biomedical demonstration, we imaged the cardiac magnetic field of a human volunteer. The key methodological advancement is the simultaneous acquisition of MCG and ECG within the passive MSR environment. While MCG sensors map the extracorporeal magnetic field generated by cardiac current distributions, the concurrent ECG provides an electrical timing reference for beat detection, classification and feature alignment. Integrating both modalities into a single session reduces physiological uncertainty and enables direct beat-to-beat comparisons, thereby enabling a direct investigation of the physiological relationship between MCG and ECG. \cite{brisinda2023clinical,fenici2005clinical,peczalski2024synchronous}

The experimental setup was designed to ensure electromagnetic compatibility between the two systems. We used galvanically isolated ECG electronics, non-magnetic leads and controlled cable routing to minimize interference. Crosstalk was evaluated through synchronized control recordings, including ECG-only and MCG-only recordings, and deliberate lead-motion tests. These benchmarks confirm that ECG-synchronous artefacts at the OPM sensors remain below the measurable noise floor, typically below 23$\,$fT$\sqrt{\mathrm{Hz}}^{-1}$, ensuring that the two modalities can be acquired simultaneously without mutual degradation.

\begin{figure}[htbp]
\centering
 \includegraphics[width=1\linewidth]{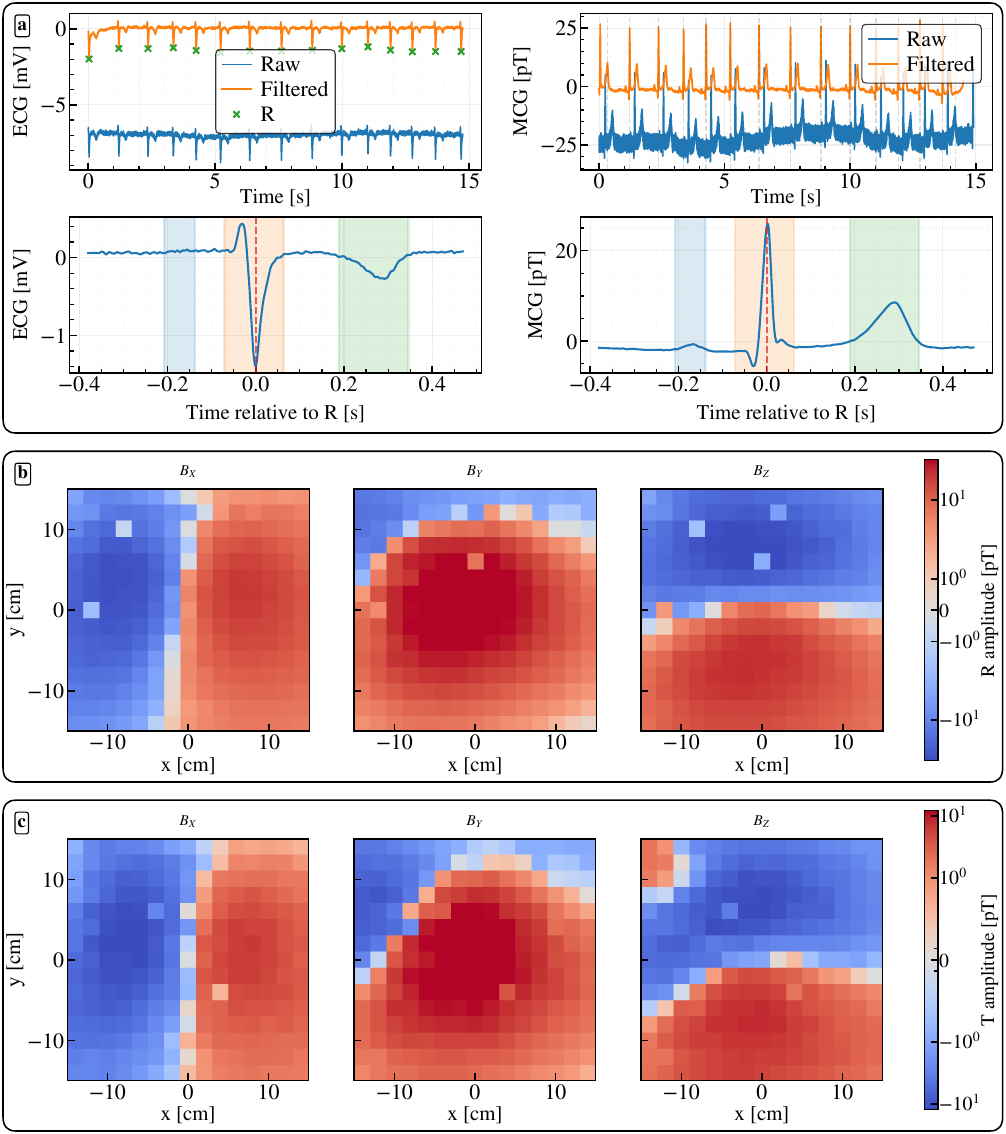}\
\caption{\textbf{Adult MCG with simultaneous ECG reference.} \textbf{a}, Simultaneously recorded ECG and MCG data, averaged over typically 15 beats. As an example, the ECG's V1 lead and a magnetic field component across the chest are shown here. The absence of measurable mutual interference enables integrated cardiac source analysis and quantitative model testing.
\textbf{b, c}, Representative MCG feature maps for the R-wave and T-wave components across the mapping grid above the middle of the chest. The spatial structure in all three field components reflects the underlying cardiac current distribution. The data from the individual measurement points is shown without any interpolation or smoothing. The small visible outliers are caused by sensor artifacts and are more visible due to the logarithmic scaling. Being unfiltered and highly resolved within the biomagnetically relevant range, this dataset serves as an ideal reference for testing models and training new neural network analyses.}
\label{fig:mcg}
\end{figure}

Measurements were performed using five OPM sensors stacked vertically with 2.5$\,$cm spacing while the mapping device scanned a 30$\,$cm $\times$ 30$\,$cm grid at a standoff of approximately 2$\,$cm from the chest. All data were acquired in the passive field environment without active cancellation. Figure~\ref{fig:mcg} displays the resulting MCG feature maps for all three field components. The spatial structures are consistent with the expected cardiac source pattern, with R-wave amplitudes reaching 10-30$\,$pT. The simultaneous ECG reference allowed systematic segmentation of cardiac cycles and robust beat detection. The preservation of raw signal characteristics and high-fidelity resolution - untainted by aggressive filtering - positions this data as a new gold-standard benchmark. Such high-quality records are essential for both rigorous biophysical model testing and the development of robust machine learning architectures for automated signal decoding.

\subsection*{Fetal magnetocardiography in a reference-quality passive environment}

A particularly sensitive application of the described magnetic environment is fMCG. Fetal MCG is recognized as the most accurate technique for diagnosing fetal arrhythmias and characterizing repolarization abnormalities, such as long QT syndrome, in utero \cite{wacker2022contribution,wacker2022jacc,strasburger2008fetal}. However, the technique remains largely confined to specialized centres because it requires either SQUID-based systems in large MSRs or careful optimization of compact, portable shields \cite{wurm2023smallscale,strasburger2008fetal}. One route to broader use is to better understand the underlying fetal and maternal magnetic signals and to develop robust source and signal-processing models that remain valid under less favourable measurement conditions.

\begin{figure}[htbp]
\centering
 \includegraphics[width=1\linewidth]{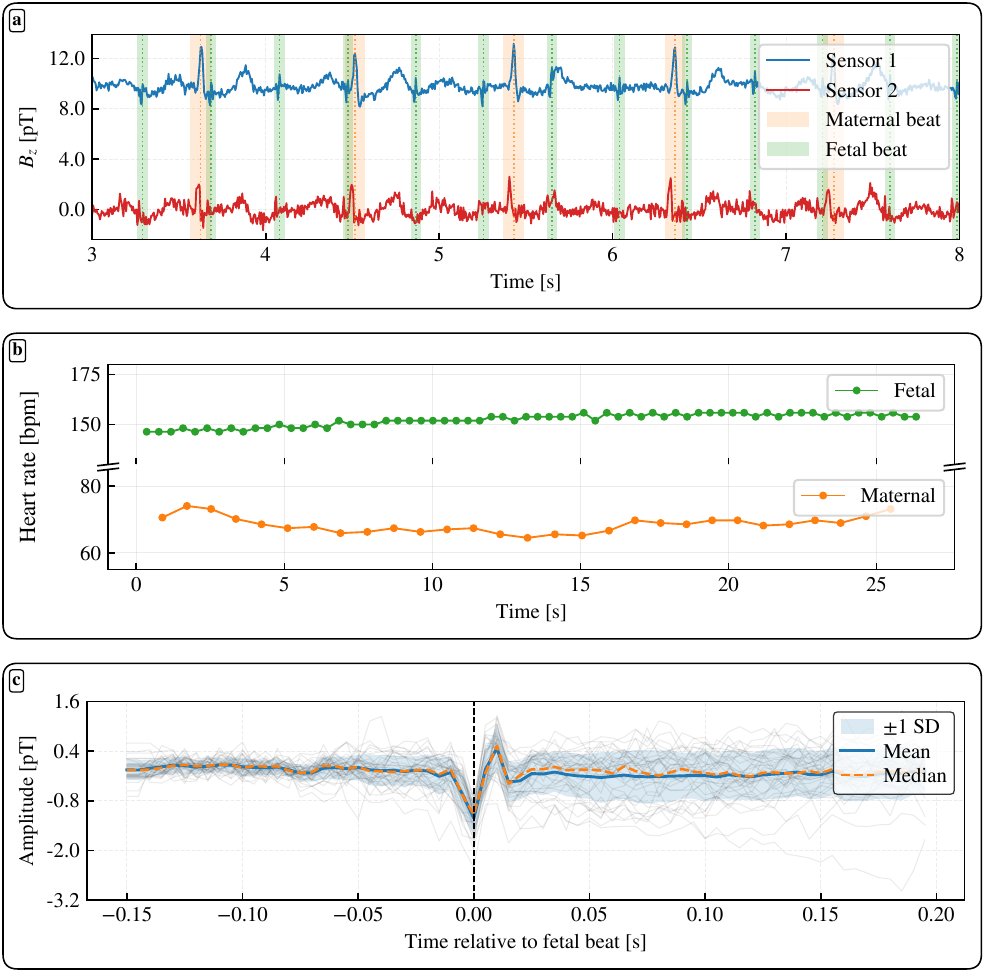}\
\caption{\textbf{Fetal MCG proof of principle at 21 weeks of gestation.} \textbf{a}, Representative minimally filtered magnetic trace showing fetal cardiac transients visible against the maternal background. \textbf{b}, Maternal and fetal heart rates over the measurement period. \textbf{c}, Extracted averaged fetal signal from a single sensor channel. To do this, the maternal template was subtracted from the data, and the fetal heartbeats were averaged. No sophisticated methods such as independent component analysis or artificial intelligence were used for the analysis. As a result, the largely unfiltered data serves as an ideal reference for better modeling and analyzing the fundamental signal.}
\label{fig:fmcg}
\end{figure}

Fetal cardiac signals are inherently small, dependent on gestational age, and typically superimposed on maternal cardiac activity, fetal motion and environmental noise. The reference-quality passive environment of this MSR provides measurement conditions that allow inspection of raw or minimally filtered signals before aggressive averaging or model-based separation is applied. As an illustrative proof of principle, we performed measurements at 21 weeks of gestation, a stage at which signal amplitudes are particularly challenging. Even at this early gestational age, fetal cardiac transients are visible in the magnetic traces (Fig.~\ref{fig:fmcg}). The figure shows the information obtained using a single sensor channel without the use of advanced algorithms such as blind source separation or spatial filtering. This provides a much deeper insight into the fundamental signal, as the resulting signal is known at a given point without model assumptions influencing the signal’s morphology. 

These data serve as a reference for developing and validating signal-processing models and diagnostic criteria for lower-performance instruments in clinical deployment.

This example demonstrates the clinical relevance of the room characterization: simultaneous MCG/ECG infrastructure allows adult cardiac maps, maternal ECG and fetal magnetic traces to be combined into a systematic validation dataset for source modelling and the training of deep learning algorithms aimed at uncovering cardiac signatures. This two-tier approach, with reference-quality characterization in this MSR followed by transfer of validated models to portable systems, provides a pathway toward establishing fMCG as a routine diagnostic tool in hospitals and fetal cardiac-care centres.

\subsection*{Standing finger-tapping MEG without active field compensation}

To demonstrate the room's capability for movement-tolerant neuroscience, we performed an OPM-MEG measurement with a subject in a standing position, a configuration that typically benefits from active field compensation \cite{boto2018moving,brookes2022opm}. The participant wore an OPM helmet array within the equilibrated MSR and performed 20$\,$s of right-hand finger tapping followed by 20
$\,$s of rest. This cycle was repeated 9 times. The sensor signals remained within their linear dynamic range throughout the experiment, despite the subject's standing posture and motion. No active field feedback or matrix-coil compensation was used.

\begin{figure}[htbp]
\centering
 \includegraphics[width=1\linewidth]{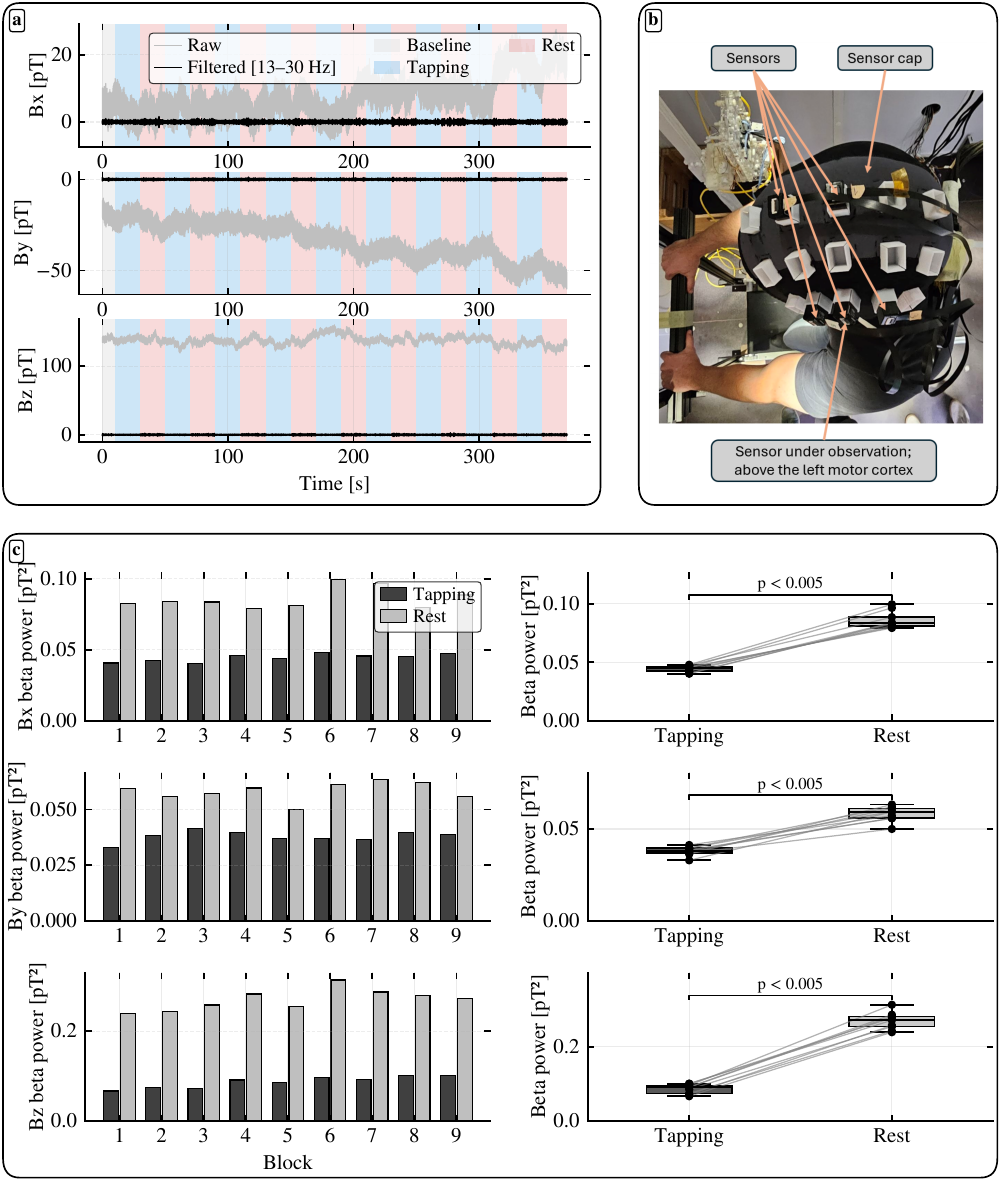}\
\caption{\textbf{OPM-MEG demonstration during standing finger tapping.} \textbf{a}, Raw field signals from a representative sensor for $B_x$, $B_y$ and $B_z$, showing modulation synchronized with tapping and rest blocks over a 360$\,$s recording. \textbf{b}, Photograph of the subject wearing the OPM helmet array inside the MSR. \textbf{c}, Beta-band, 13--30$\,$Hz, power analysis per block and tapping-versus-rest comparison. Beta desynchronization is observed across all channels. }
\label{fig:meg}
\end{figure}

The analysis focused on task-related modulation of beta-band power, 13--30$\,$Hz, a canonical signature of motor cortex engagement \cite{pfurtscheller1999erd,cheyne2013meg}. We observed beta event-related desynchronization during movement compared with rest in all three sensor axes, with $p<0.005$ in a Wilcoxon signed-rank test across task and rest blocks (Fig.~\ref{fig:meg}). This proof-of-principle demonstration validates that the passive environment is sufficiently stable and low-gradient to resolve established sensorimotor MEG contrasts in a standing participant without active compensation.

\subsection*{Ultra-low-field nuclear spin precession}

The passively generated residual field and reproducible correction strategy enable nuclear spin precession in a regime that is largely inaccessible to conventional low-field nuclear magnetic resonance. We demonstrate this through free induction decay measurements of polarized $^3$He nuclear spins, detected by the OPM array at holding fields ranging from 50$\,$nT down to approximately 100$\,$pT.

A primary challenge in precision spin precession is transverse relaxation induced by magnetic-field gradients.
In the motional-narrowing regime, the contribution of field gradients to the relaxation rate, $1/T_2^{\mathrm{ind}}$, is determined by the interplay of diffusion and magnetic field spatial variations \cite{cates1988relaxation}.
At the passively achieved MSR level, $|\nabla B|\approx 100$$\,$pT/m, the calculated gradient-induced coherence time for a typical cylindrical cell with a diameter and length of approximately 5 cm is $T_2^{\mathrm{ind}}\approx 421$$\,$h. With static corrections, $|\nabla B|\approx 1$--10$\,$pT/m, this theoretical limit extends to between 42,000 and more than four million hours. These values are significantly higher than the best demonstrated intrinsic $^3$He cell $T_2$ values, $\sim 100$--150$\,$h, which are typically limited by wall relaxation \cite{heil1995longt2}. Consequently, this MSR environment effectively eliminates the background gradient as a dominant source of decoherence, shifting the experimental focus to cell properties and holding-field uniformity. It is interesting to note that the magnetization of the $^3$He cell is much higher than the NMR field: here the spins are strongly coupled, which is a new and previously inaccessible regime with effects on coherence and thus the macroscopic behavior of such a system.

\begin{figure}[htbp]
\centering
 \includegraphics[width=1\linewidth]{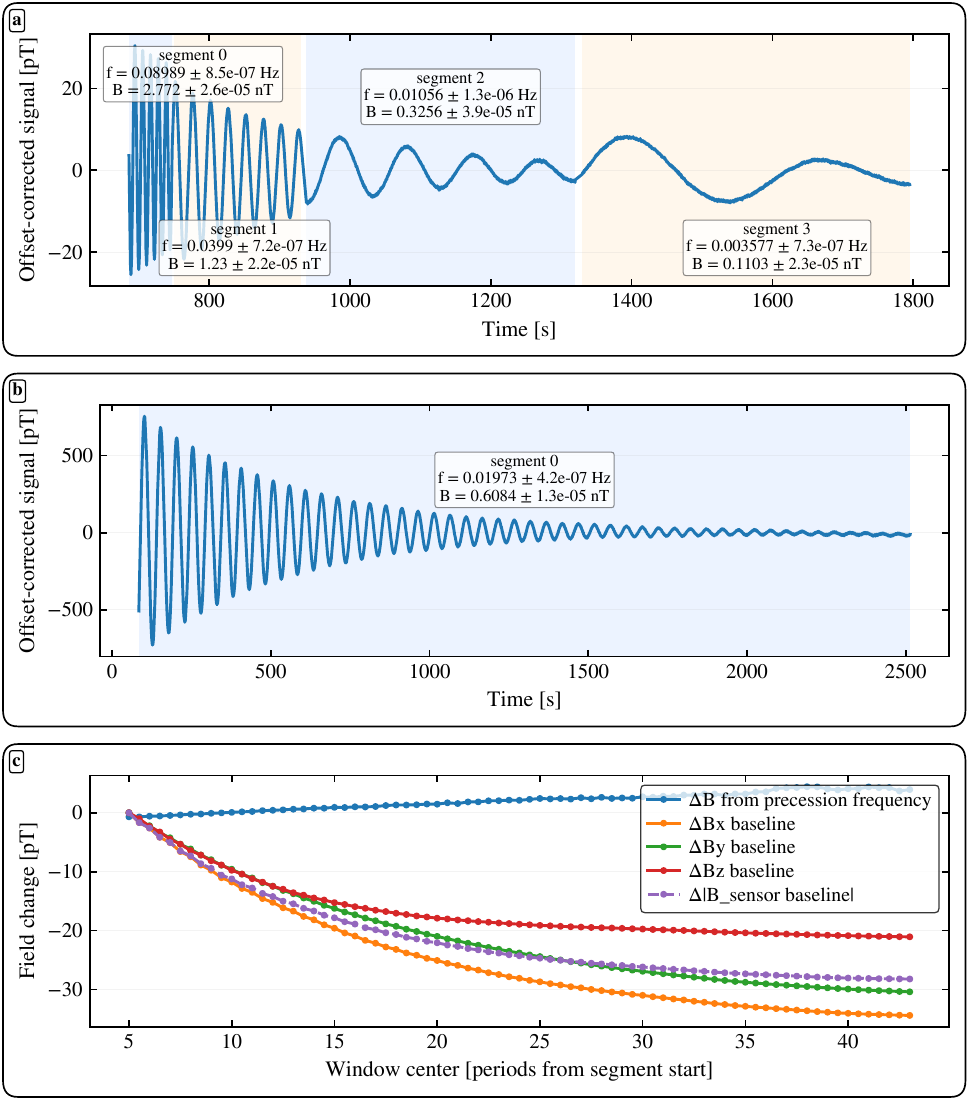}\
\caption{\textbf{Ultra-low-field $^3$He precession and long-term stability.} \textbf{a}, Extracted spin-precession segments from a representative $^3$He free-induction decay at decreasing holding fields. The Larmor frequency $f_L$ is tracked from $\sim 90$$\,$mHz, corresponding to $B_0\approx 2.8$$\,$nT, down to $\sim 3.6$$\,$mHz, corresponding to $B_0\approx 0.11$$\,$nT. 
\textbf{b}, Long-term stability demonstration: a continuous 2,500$\,$s recording of a $^3$He precession signal at $B_0\approx 608$$\,$pT, corresponding to $f_L\approx 20$$\,$mHz. The high reproducibility of the passive field allows precise frequency extraction and separation of true magnetic signals from sensor-intrinsic drifts.
\textbf{c},Differences in sensor drift during a time-series measurement of $^3$He precession. One can see the difference between the temporal field variation, which was determined by the sensor baseline alone, and the field variation resulting from the helium precession frequency. As can be seen, the sensor reading changes by about 30 pT over the time period, while the field in the cell changes by only about 2 pT, confirming the assumption that, in the system comprising the sensor and MSR, the drift in the measured magnetic field is dominated by the sensor’s intrinsic drift. }
\label{fig:he3}
\end{figure}

A defining feature of this environment is the ability to resolve precession at extremely low frequencies. As shown in Fig.~\ref{fig:he3}a, we tracked $^3$He free-induction-decay segments at holding fields as low as 0.11$\,$nT, corresponding to a Larmor frequency of approximately 3.6$\,$mHz. In this quasi-DC regime, where a single oscillation period may span several hundred seconds, the measurement is exceptionally sensitive to sensor baselines and environmental stability. To demonstrate suitability for long-duration precision experiments, we recorded a continuous spin-precession signal at $\sim 20$$\,$mHz over 2,500$\,$s (Fig.~\ref{fig:he3}b). The stable passive background allows direct fitting of the decaying sinusoidal signal, separating the oscillatory nuclear frequency from slow OPM baseline drifts. This capability is a prerequisite for next-generation EDM searches, where sensitivity scales with coherence time and the stability of the magnetic environment over long observation windows.

\section*{Discussion}

The results presented here establish this MSR as a qualitatively distinct magnetic environment. The core innovation is not merely the low residual field, but its deterministic reproducibility. By demonstrating picotesla-scale reproducibility of the residual-field pattern, with component-wise differences at the few-picotesla level and map-level standard deviations of order 10$\,$pT, we transform the residual field from an uncontrolled nuisance into a measurable, repeatable and correctable platform. The current measurements remain primarily limited by sensor-intrinsic noise and mechanical uncertainties, such as the spatial positioning of the sensors and the sample, rather than the magnetic environment itself. This indicates that the MSR provides a surplus of shielding performance, establishing a future-proof platform where future scientific sensitivity will be driven by sensor innovation and mechanical refinement rather than environmental constraints.\\

% \subsection*{Implications for biomagnetism}

For OPM-based neuroimaging and cardiology, the primary implication is the elimination of active field compensation as a prerequisite for the measurements demonstrated here. Current wearable MEG systems often rely on active feedback loops that can introduce additional noise from current sources and reference sensors \cite{boto2018moving,holmes2022lightweight,holmes2023ambulatory,brookes2022opm}. Our demonstration of beta-band desynchronization in a standing subject without active feedback confirms that a sufficiently stable passive environment can simplify instrumentation while preserving a 23$\,$fT$\sqrt{\mathrm{Hz}}^{-1}$ sensor noise floor.

The simultaneous MCG and ECG measurements show how this environment can generate systematic validation datasets. By acquiring electrical timing references alongside magnetic maps, the setup bridges engineering performance and clinical source modelling. 
Conventional measurements in magnetically shielded environments often necessitate aggressive digital signal processing, such as notch filtering for power-line interference or other noise sources in the environment or other model-dependent algorithms that can alter the temporal and spatial structure of the data. While effective for signal visualization, such methods inevitably introduce phase distortions and can attenuate transient, high-frequency physiological features.
The availability of high-bandwidth, unfiltered data within this MSR enables several advanced modeling strategies:

\begin{itemize}
    \item \textbf{Higher-Order Multipole Expansion:} The spatial homogeneity and low background noise facilitate the use of multipole models beyond the standard equivalent current dipole (ECD). Stable reconstruction of quadrupole and octupole moments is now feasible, providing deeper insights into complex excitation patterns, particularly in fMCG where the signal-to-noise ratio is traditionally limiting.
    \item \textbf{High-Frequency Oscillation (HFO) Analysis:} The pristine noise floor at higher frequencies opens a window for analyzing HFOs ($>200$~Hz). These are increasingly recognized as critical biomarkers for epileptogenic zones in MEG and early myocardial ischemia in MCG, but are typically masked by technical noise in standard facilities.
    \item \textbf{Validation of Individualized Conductivity Models:} The high reproducibility of the residual field allows for a direct comparison between raw experimental data and numerical finite-element method (FEM) simulations. By minimizing pre-processing artifacts, errors in the forward model can be isolated and corrected, paving the way for patient-specific precision medicine in biomagnetic imaging.
    \item \textbf{Portable use in  Applications:} For fMCG, the same stability enables minimally filtered early-gestation magnetic traces that can serve as reference data for algorithms intended for smaller or portable systems.
\end{itemize}

% \subsection*{Reproducible gradients and Berry-phase suppression}

The combination of a reproducible residual field and ultra-low-field $^3$He precession motivates a different strategy for EDM searches. In trapped-particle or gas-cell experiments, particles moving in an electric field experience the motional magnetic field. In the presence of transverse magnetic-field gradients, this field can generate a geometric-phase frequency shift that is linear in the electric field and therefore indistinguishable from a true EDM signal if not controlled \cite{chupp2019edm,pendlebury2004geometric,harris2006dipole}.

The relevant advantage of the present MSR is not only that the raw gradients are small, but that the residual gradient tensor is reproducible after magnetic equilibration. This enables a mode of operation in which the magnetic background is measured before an EDM run and treated as a calibrated boundary condition rather than an unknown initial state. The measured residual-field map can then be used to align the experiment, choose the operating volume, apply static correction fields and include the remaining gradients directly in Monte-Carlo trajectory simulations and the systematic-error budget \cite{pendlebury2004geometric,harris2006dipole}.

This does not eliminate geometric phases automatically. Instead, it changes the problem from suppressing an unknown, fluctuating field to correcting and modelling a deterministic field pattern. In this sense, a reproducible passive MSR provides a route toward Berry-phase-suppressed EDM searches in which the residual-gradient contribution is measured, minimized by design and independently verified across repeated equilibration cycles.\\

% \subsection*{Disentangling sensor drift, room-field drift and demagnetizing-field shifts}

The simultaneous use of OPM readout and nuclear spin precession provides a diagnostic tool for separating sensor-intrinsic drift from true magnetic-field drift in the shielded room. The OPM measures the local magnetic field at the sensor head and is therefore sensitive to laser, heater, electronics and polarimetry drifts. In contrast, the $^3$He Larmor frequency measures the magnetic field experienced by the nuclear spins inside the cell. A drift in the OPM baseline without a corresponding change in the nuclear precession frequency identifies a sensor-specific drift, whereas a true change of the holding field shifts the nuclear frequency according to $\delta f=\gamma\delta B/(2\pi)$ \cite{cates1988relaxation,chupp2019edm,heil1995longt2}. In this sense, the precessing noble-gas sample acts as an independent magnetic clock against which OPM baseline drift and room-field drift can be compared.

This interpretation requires one important qualification: the nuclear precession frequency is not a pure monitor of the external room field. The polarized noble-gas sample generates its own magnetic field, and this self-field can shift the Larmor frequency through demagnetizing-field effects that depend on the cell geometry and magnetization distribution. In the simplest description, the internal self-field of a uniformly magnetized sample is described by a demagnetization tensor. For a highly symmetric cell, this contribution can be calculated or strongly reduced by symmetry. A real cell with a one-sided filling stem or tube breaks this symmetry and can generate a non-zero dipolar field averaged over the main cell volume, even if its contribution to the external sensor signal is small \cite{harris2006dipole,osborn1945demag,terrano2019frequency,limes2019dipolar}.

For drift separation, the relevant quantity is not only the absolute frequency offset caused by such a stem-induced field, but how this contribution changes during the observation time. A constant self-field contribution can be absorbed into the fitted Larmor frequency of a given run. A relaxing stem magnetization, however, produces a time-dependent frequency contribution that can mimic slow room-field drift if it is not included in the model. For a longitudinal stem magnetization, the apparent frequency contribution can be written as
\[
\delta f_{\mathrm{stem}}(t)
=
\frac{\gamma}{2\pi}
\left\langle B_{\mathrm{stem},z}(\mathbf{r},t)\right\rangle_{\mathrm{cell}}
=
\delta f_0 \exp(-t/T_{\mathrm{stem}}),
\]
where $T_{\mathrm{stem}}$ is an effective relaxation time of the stem magnetization. This time constant need not be identical to the relaxation time of the main cell volume because the stem has a different surface-to-volume ratio, diffusion geometry and polarization history. A short $T_{\mathrm{stem}}$ produces an early transient; a long $T_{\mathrm{stem}}$ produces an almost constant offset over a short fitting window; and an intermediate $T_{\mathrm{stem}}$ produces the largest ambiguity with slow room-field drift. Transverse precessing magnetization can additionally generate Ramsey--Bloch--Siegert-type self-shifts, which scale with the square of the oscillating magnetization field and therefore decay approximately as $\exp(-2t/T_2^*)$ \cite{terrano2019frequency,limes2019dipolar}.

The reproducibility of the present MSR is essential for separating these contributions. A true room-field drift should be common to repeated measurements and consistent with independent field maps or reference magnetometer channels. An OPM baseline drift appears in the sensor signal but not in the fitted nuclear phase. A demagnetizing-field or stem-induced self-shift follows the polarization history and cell geometry, and should change predictably with cell orientation, stem orientation, initial polarization and relaxation time. The practical diagnostic hierarchy is therefore to map the passive magnetic background, monitor OPM baselines, fit the nuclear phase with possible relaxation-dependent self-shift terms, and validate the interpretation using changed cell or stem orientations, different initial polarizations or symmetric stemless cell geometries. The role of the reproducible MSR is to turn the external magnetic background into a measured and repeatable quantity, so that remaining relaxation-correlated frequency changes can be assigned to sensor drift, room-field drift or cell-intrinsic demagnetizing-field shifts rather than being conflated.\\

% \section*{Application to sub-hertz dark matter detection}
To demonstrate the foundational physics impact of the constructed MSR, we project its empirical performance onto the exclusion limits of the axion-nucleon coupling constant ($g_{aNN}$) within the canonical ultralight dark matter haloscope framework~\cite{garcon_cosmic_2018}. When invoking quantum spin ensembles for detection, the coherent dark matter field acts as an equivalent pseudo-magnetic field oscillating at its Compton frequency~\cite{garcon_cosmic_2018}. Under identical sensor configurations, the macroscopic boundaries of the MSR dictate the ultimate experimental sensitivity floor via a joint spatio-temporal constraint relation:

\begin{equation}
g_{aNN}^{\text{limit}} \propto \frac{1}{\Phi(\nabla B)} \cdot \sqrt{S_B} \cdot \sqrt{\frac{\gamma_N \left(\frac{\partial B}{\partial t}\right)}{\Delta\nu_{\text{DM}}}}
\end{equation}

where $\Phi(\nabla B)$ is the non-linear dephasing penalty driven by the spatial magnetic field gradient, $S_B$ is the background magnetic noise power spectral density, $\partial B/\partial t$ is the temporal drift rate, $\gamma_N$ the nuclear gyromagetic ratio, and $\Delta\nu_{\text{DM}}$ is the natural linewidth of the virialized dark matter halo ($Q \sim 10^6$).

For comparison, we use the values from a world-leading shielded room \cite{ayres2024ultralow, ayres2022n2edm}. Substituting our MSR's core parameters ($\nabla B \approx 100\,\text{pT/m}$ and $\partial B/\partial t \approx 1\,\text{pT/hour} \approx 2.78 \times 10^{-16}\,\text{T/s}$) into the analytical coupling mapping, our slightly higher spatial inhomogeneity introduces a minor $1.67$-fold dephasing penalty. Crucially, however, this spatial concession is heavily overcompensated by a $14.1$-fold sensitivity enhancement in the Fourier domain, directly unlocked by the $200$-fold suppression of the low-frequency temporal drift rate. Consequently, for targeted narrow-line searches where continuous coherent integration is bounded by field wander out of the dark matter spectral slit, our MSR projects an $8.5$-fold deeper exclusion depth ($g_{aNN}^{\text{limit}}$) than current world-leading infrastructure.
\\
\\
In summary, the facility described here represents a shift in strategy for high-precision magnetic measurements: from active suppression of an unknown and fluctuating environment to precision calibration of a deterministic and reproducible one. By demonstrating that the magnetic state of a multi-layer MSR can be treated as a repeatable function of its equilibration protocol, we provide a pathway to mitigating persistent systematic errors in both medicine and physics. For biomagnetism, this enables complex neuroimaging and fetal diagnostics without the noise penalties of active compensation. For fundamental physics, it allows a new generation of EDM searches and spin-precession experiments in which the residual gradient is no longer a limiting nuisance, but a known, modelled and corrected parameter. As quantum sensors continue to approach their theoretical limits, the availability of such a metrological-grade passive environment will be essential for translating picotesla-scale stability into new scientific discoveries.

\section*{Methods}

\subsection*{MSR construction and layer assembly}

The MSR is a rectangular enclosure with external dimensions of $3550\times3550\times3400$$\,$mm$^3$ and a usable internal volume of $2200\times2200\times2100$$\,$mm$^3$. The passive shielding consists of four high-permeability Magnifer layers with thicknesses of 2, 4, 4 and 3$\,$mm, each electrically isolated to minimize eddy-current coupling between layers. The innermost high-permeability layer is assembled under controlled mechanical conditions: panels are handled with padded fixtures to prevent point stress concentrations and panel-to-panel overlap joints use controlled contact pressure. An additional aluminium shell provides radio-frequency shielding and eddy-current damping for higher-frequency interference \cite{bork2000bmsr2,thiel2007demagnetization,kadenShielding}.

The infrastructure includes a specialized door system: the inner two layers are integrated into a sliding door, while the outer layers use a swing door. This geometry ensures that the innermost shielding layer is not directly exposed to the Earth's magnetic field during access, maintaining the equilibrated state. All interfaces, including air exchange, gas lines and cabling, are routed through waveguide-style penetrations optimized to maintain shielding integrity. Lighting is provided via optical fibres, and a projector system operates through a waveguide to eliminate internal electronic noise.

\subsection*{Magnetic equilibration procedure}

The magnetic equilibration process is fundamental to achieving a stable low-energy state of the high-permeability layers \cite{altarev2014msr,altarev2015minimizing,allmendinger2023degaussing,sun2021limits,thiel2007demagnetization}. The procedure uses dedicated coil sets wound around the innermost and second-innermost shielding layers. We employ a sequential protocol: the innermost layer is equilibrated first, followed by the second-innermost layer and concluding with a final equilibration of the innermost layer to define the internal environment.

The hardware setup consists of an automated switching unit with DC relays, controlled by an Arduino Mega and a LabJack U6. A secondary control box manages the power supply for the relays and includes switching circuitry to ensure that the amplifier signal returns to zero without current jumps. The degaussing signal is a decaying sine wave, sampled at 10$\,$kHz with an NI USB-6343 16-bit digital-to-analogue converter, and modulated with a cosine window. To eliminate DC offset from the electronics, the amplified signal from a Rohrer current amplifier passes through a transformer before being routed to the coils. The system also performs automatic resistance checks of the selected coil configuration before each run.

\subsection*{Non-magnetic robotic mapper}

The spatial field distribution was characterized using a custom-built automated three-dimensional mapping system. To minimize magnetic interference, all stepper motors and control electronics are positioned outside the MSR. Mechanical motion is transmitted to the interior carriage via non-magnetic, non-conductive fiberglass rods and inelastic cords. The system comprises two horizontal axes for $xy$ positioning and a vertical axis for height adjustment, controlled by an Arduino Nano. This setup allows reproducible sensor movement with uncertainties in the millimetre range without perturbing the ultra-low-field environment and can typically remain permanently installed in the room and be moved to the side.

\subsection*{Magnetometers and calibration}

Characterization measurements were performed using three-axis fluxgate sensors, Bartington Mag-03, with a specified noise of 6$\,$pT/$\sqrt{\mathrm{Hz}}$, and triaxial QZFM OPMs, QuSpin, with a specified noise of 23$\,$fT$\sqrt{\mathrm{Hz}}^{-1}$ \cite{quspinQZFM,bartingtonMag03}. To eliminate systematic sensor offsets, we developed a non-magnetic rotation holder. Constructed from 3D-printed parts and acrylic glass, the holder features two orthogonal turntables. To maintain a personnel-free environment during calibration, rotation is driven by pneumatic actuators and rubber bands, allowing remote-controlled 180$^\circ$ flips of the sensors.

This holder is mounted directly onto the robotic mapper. This integration allows offset determination and field mapping in a single automated workflow: by measuring the field in two opposite orientations at each grid point, we calculate the absolute magnetic field and the sensor offset, ensuring that the reported maps are not dominated by long-term sensor drift or electronic offsets.

\subsection*{Shielding factor measurement}

The passive shielding factor was determined by applying a known external magnetic field and measuring the attenuated field inside the MSR with an OPM. External coils generated a homogeneous excitation field. The shielding factor is calculated as the ratio of external field amplitude to internal field amplitude, $SF=B_{\mathrm{ext}}/B_{\mathrm{int}}$. Measurements were performed across 0.001--100$\,$Hz to characterize both flux-shunting by the Magnifer layers and eddy-current shielding by the aluminium shell.

\subsection*{Residual field and reproducibility analysis}

The reproducibility of the magnetic state was evaluated by performing independent equilibration cycles. After each cycle, the robotic mapper acquired a full field map of the central volume with initial and final sensor-offset calibration. These maps were subtracted from one another to determine cycle-to-cycle stability. This analysis quantifies the deterministic nature of remanent magnetization, demonstrating that the residual field is not a random distribution but a repeatable function of the equilibration protocol.

The measurements shown in the Results were recorded using an OPM. Additional fluxgate measurements were acquired to verify the results and were consistent with the OPM data within the fluxgate uncertainty.

\subsection*{Noise and Allan deviation}

The temporal stability and noise floor of the MSR were analysed using OPM and fluxgate data. To identify power-line interference and other technical noise sources, linear spectral density (LSD) estimates were computed using Welch's method. Prior to spectral estimation, the global mean was subtracted from the data. We applied a Nuttall window with a segment length of 10 s and a 50\% overlap, utilizing linear detrending within each segment. The LSD was then obtained by taking the square root of the one-sided power spectral density. To characterize long-term stability, we calculated the Allan deviation $\sigma_y(\tau)$ \cite{allan1966statistics}. For the fluxgate sensors, the Allan deviation reached a minimum of approximately 4$\,$pT for integration times between 30 and 200$\,$s, which informed our choice of a 30$\,$s measurement duration per mapping point. For the $^3$He experiments, the noise floor was evaluated specifically in the sub-Hz regime to ensure that frequency fitting is not limited by environmental fluctuations.

\subsection*{Simultaneous MCG and ECG acquisition}

Adult MCG was recorded using OPM sensors mapped in a grid above the chest. Simultaneously, ECG was acquired using a galvanically isolated analogue-to-digital converter and non-magnetic electrodes and connectors. A shared synchronization signal was used to align the acquisition systems. Crosstalk was evaluated using control runs, including ECG-only and MCG-only recordings, and deliberate lead-motion tests to establish an upper bound on synchronous artefacts. Beat detection and feature mapping used the ECG as the primary timing reference to avoid systematic errors caused by the spatially variable MCG waveform.

For each map position, the MCG and ECG traces were loaded together with their recorded time axes and metadata. If required, the ECG time axis was reconstructed from the effective sampling frequency rather than the nominal sampling rate, using
\[
    t_{\mathrm{ECG}} = t_0 + \frac{n}{f_{\mathrm{s,ECG}}},
\]
with \(f_{\mathrm{s,ECG}} = 1017\,\mathrm{Hz}\) for datasets in which the LabJack sampling rate deviated from the nominal \(1000\,\mathrm{Hz}\). The MCG time axis was taken from the NI acquisition metadata, typically sampled at \(f_{\mathrm{s,MCG}} = 1500\,\mathrm{Hz}\).

The ECG traces were band-pass filtered between \(0.3\,\mathrm{Hz}\) and \(95\,\mathrm{Hz}\). The MCG traces were filtered between \(0.5\,\mathrm{Hz}\) and \(80\,\mathrm{Hz}\). All filtering and subsequent averaging steps were performed separately for each map position. To avoid selecting ECG channels dominated by motion artifacts or non-cardiac transients, the ECG lead used for beat detection was selected once at the first usable map position after an automatic channel-quality assessment and was then kept fixed for the remaining positions. ECG R-peaks were detected using a hybrid detector combining neural-network-based ECG segmentation with a heuristic fallback based on peak prominence, inter-beat spacing and physiological plausibility.

Beat-averaged ECG and MCG waveforms were computed by extracting fixed windows around the detected ECG R-peaks. The ECG R-peak times were mapped to the MCG time axis using the synchronization information available in the original acquisition pipeline. These mapped times were then used as seed times for the MCG beat windows, so that the MCG averages were referenced to the ECG-defined cardiac cycle rather than to position-dependent MCG morphology. For each map position, this yielded beat-averaged MCG traces for all sensor components and a corresponding beat-averaged ECG trace.

ECG feature positions, including \(P\), \(Q\), \(R\), \(S\), \(T\), and onset/offset markers, were reviewed interactively at the first valid map position. The resulting feature times were stored as offsets relative to the ECG R-peak and subsequently applied to all other map positions. This ensured that feature-based MCG maps were evaluated at consistent ECG-defined cardiac time points. Points with insufficient ECG quality, erroneous R-peak detection, or poor ECG--MCG alignment were flagged during preprocessing and could be corrected in a post-hoc interactive review, where individual false beats could be removed and a manual ECG-to-MCG time shift could be applied before recomputing the local averages.

\subsection*{Fetal MCG measurement}

Fetal MCG measurements were performed with the same acquisition framework as the adult MCG measurements, with the sensors placed over the pregnant participant's abdomen. In the 21-week gestational example, individual fetal complexes were identified in minimally filtered data. \\
The fMCG data consisted of five \(B_z\) sensor channels sampled at \(200\,\text{Hz}\). After import, the longest continuous segment without
missing values was selected, linearly detrended, and zero-phase band-pass
filtered between \(1\,\text{Hz}\) and \(90\,\text{Hz}\). Maternal and fetal beats were manually annotated in an interactive viewer using the filtered sensor traces. Beat positions were refined at sample-level precision and exported as maternal and fetal trigger times.

For visualization, the data were additionally filtered in maternal-dominant \(0.8\,\text{Hz}- 2.0\,\text{Hz}\) and fetal-dominant \(2.0\,\text{Hz}-4.0\,\text{Hz}\) frequency bands. Beat-triggered averages were computed by extracting epochs around the manual triggers, applying baseline correction and local temporal realignment. Mean, median, and single-epoch overlays were used to assess signal consistency.
Maternal interference was reduced using manually triggered template subtraction. For each selected sensor, a maternal template was computed as the median of baseline-corrected maternal epochs. Several subtraction variants were compared, ranging from fixed-template subtraction to adaptive subtraction with local amplitude and timing correction. In the adaptive model, each maternal epoch was fitted as
\[
    y(t) \approx a\,T(t-\Delta t) + b + c\,t ,
\]
where \(T\) is the maternal template, \(a\) is a local scale factor,
\(\Delta t\) is a small timing shift, and \(b+c\,t\) accounts for local baseline and trend. Only the fitted maternal component \(a\,T(t-\Delta t)\) was subtracted, using a tapered window to avoid edge artifacts. Subtraction variants were compared using original-versus-processed time traces, fetal beat-triggered averages, spectral summaries, and quantitative metrics including residual maternal RMS and fetal average peak-to-peak amplitude.

\subsection*{OPM-MEG finger-tapping analysis}

The MEG demonstration involved a standing participant wearing an OPM helmet array. The experiment consisted of alternating 20$\,$s blocks of right-hand finger tapping and rest. Data were analysed in the beta band, 13--30$\,$Hz, to identify event-related desynchronization \cite{pfurtscheller1999erd,cheyne2013meg}. Statistical significance was determined using a non-parametric Wilcoxon signed-rank test comparing tapping and rest blocks. Because this proof-of-principle measurement was performed in a single participant, p-values are interpreted as within-participant evidence of task modulation rather than population-level inference.

\subsection*{Low-field $^3$He precession and frequency fitting}

$^3$He was polarized using spin-exchange optical pumping with rubidium. The polarized $^3$He cell was placed in the central volume of the MSR, and holding fields of a few nanotesla down to a few hundred picotesla were applied using coils on the mu-metall in a Magic Box configuration. To observe Larmor precession, spin flips were implemented using two methods. In the first approach, an adiabatic $\pi/2$ flip was performed by applying an oscillating magnetic field perpendicular to the main holding field. In the second approach, a non-adiabatic flip was performed by abruptly switching off a high-amplitude holding field, leaving a residual orthogonal field component as the new quantization axis around which the spins precessed. The precession signal was detected by the OPM array.

The time-series data were fitted using a decaying sinusoidal model,
\[
B(t)
=
A\exp(-t/T_2)\sin(2\pi f t+\phi)
+
c_0+c_1t+c_2t^2,
\]
where the polynomial terms account for low-order baseline drifts of the OPM readout. Frequency uncertainty was evaluated using the covariance matrix and residual diagnostics. This fit separates the oscillatory nuclear precession frequency from slowly varying OPM baselines. A true drift of the holding field appears as a change in the fitted nuclear frequency, whereas an OPM baseline drift primarily appears in the polynomial background terms or in the non-oscillatory sensor channel.

\subsection*{Magnetostatic estimate of stem-induced noble-gas frequency drifts}

To estimate the possible size of cell-intrinsic frequency changes that could complicate the interpretation of long precession records, we calculated the magnetic field generated by polarized gas in a one-sided filling stem or tube and averaged its axial component over the main cell volume. This estimate was used to assess the systematic scale of demagnetizing-field and stem-induced self-shifts; it was not used as a correction to the measured MSR field drift.

The polarized gas was treated as a magnetized volume with magnetization
\[
M=nP\mu ,
\]
where $n$ is the noble-gas number density, $P$ the nuclear polarization and $\mu$ the nuclear magnetic moment. The main cell was approximated as a cylinder with radius $R=2.5\,\mathrm{cm}$ and length $L=5.0\,\mathrm{cm}$. The one-sided filling stem was approximated as a cylinder with diameter $d_{\mathrm{stem}}=3\,\mathrm{mm}$ and length $l_{\mathrm{stem}}=5.0\,\mathrm{cm}$. The corresponding stem-to-cell volume ratio is
\[
\frac{V_{\mathrm{stem}}}{V_{\mathrm{cell}}}
=
\frac{(d_{\mathrm{stem}}/2)^2 l_{\mathrm{stem}}}{R^2L}
\approx
3.6\times10^{-3}.
\]
Although this volume fraction is small, the asymmetric stem magnetization produces a non-zero axial dipolar field averaged over the main cell volume.

The axial field component of the magnetized stem was averaged over the main cell volume,
\[
\left\langle B_{\mathrm{stem},z}\right\rangle_{\mathrm{cell}}
=
\frac{1}{V_{\mathrm{cell}}}
\int_{\mathrm{cell}}
B_{\mathrm{stem},z}(\mathbf{r})\,\mathrm{d}^3r .
\]
For the geometry used here, the magnetostatic estimate gives approximately
\[
\left\langle B_{\mathrm{stem},z}\right\rangle_{\mathrm{cell}}
\approx
734\,\mathrm{pT}
\left(\frac{M}{1\,\mathrm{A/m}}\right).
\]
The corresponding apparent frequency contribution is
\[
\delta f_0
=
\frac{\gamma}{2\pi}
\left\langle B_{\mathrm{stem},z}\right\rangle_{\mathrm{cell}}.
\]
For $^3$He at $1\,\mathrm{bar}$ and polarization $P=80\%$, the larger nuclear magnetic moment gives $M\approx2.1\times10^{-1}\,\mathrm{A/m}$, a volume-averaged stem field of approximately $154\,\mathrm{pT}$ and an initial apparent frequency contribution of
\[
\delta f_0(^3\mathrm{He})
\approx
5.0\,\mathrm{mHz}.
\]
These values are geometry-specific scale estimates for a longitudinally magnetized one-sided stem.

For drift interpretation, we modeled the longitudinal stem contribution as an exponentially relaxing frequency term,
\[
\delta f_{\mathrm{stem}}(t)
=
\delta f_0 \exp(-t/T_{\mathrm{stem}}),
\]
where $T_{\mathrm{stem}}$ is the effective relaxation time of the stem magnetization. The change of the stem-induced frequency contribution over an observation time $T_{\mathrm{obs}}$ is
\[
\Delta f_{\mathrm{stem}}(T_{\mathrm{obs}})
=
\delta f_0
\left[
1-\exp(-T_{\mathrm{obs}}/T_{\mathrm{stem}})
\right],
\]
whereas the time-averaged contribution to a simple frequency estimate over the same interval is
\[
\left\langle \delta f_{\mathrm{stem}}\right\rangle
=
\delta f_0
\frac{T_{\mathrm{stem}}}{T_{\mathrm{obs}}}
\left[
1-\exp(-T_{\mathrm{obs}}/T_{\mathrm{stem}})
\right].
\]
The accumulated phase from this exponentially decaying contribution is
\[
\phi(T_{\mathrm{obs}})
=
2\pi
\int_0^{T_{\mathrm{obs}}}
\delta f_0\exp(-t/T_{\mathrm{stem}})\,\mathrm{d}t
=
2\pi \delta f_0 T_{\mathrm{stem}}
\left[
1-\exp(-T_{\mathrm{obs}}/T_{\mathrm{stem}})
\right].
\]

The initial value $\delta f_0$ therefore sets the scale of the cell-intrinsic self-shift, while $T_{\mathrm{stem}}$ determines whether this contribution appears mainly as a constant frequency offset or as an observable drift during the measurement. For $T_{\mathrm{stem}}\gg T_{\mathrm{obs}}$, the term is nearly constant over the fitting window and is largely absorbed into the fitted frequency. For $T_{\mathrm{stem}} \leq  T_{\mathrm{obs}}$, it produces a relaxation-correlated frequency change that can mimic slow room-field drift. Because the stem has a larger surface-to-volume ratio than the main cell and may differ in rubidium exposure and diffusion geometry, $T_{\mathrm{stem}}$ was treated as an effective nuisance parameter rather than fixed to the main-cell relaxation time. Assuming $T_{\mathrm{obs}}$ = 1000 s, the example of 3He with $ \delta f_0 \simeq 5.0$$\,$mHz yields average contributions of approximately 0.025$\,$mHz, 0.050 mHz, 0.25 mHz, and 0.50 mHz for  $T_{\mathrm{stem}}$ = 50 s, 100 s, 500 s, and 1000 s, respectively.

\section*{Ethics statement}

Human measurements were performed by the Department of Congenital Heart Disease, German Heart Center, TUM University Hospital, TUM School of Medicine and Health and the Department of Physics Chair of Precision Measurements at Extreme Conditions, TUM School of Natural Sciences. The protocol for this explorative study was approved by our institutional review board.

\section*{Data availability}
An anonymized data set can be made available upon request. Access is only granted to academic institutions and after signing a data share agreement.

\section*{Code availability}
The used code can be made available upon request. Access is only granted to academic institutions and after signing a data share agreement.

\section*{Acknowledgements}
This work was supported by the Deutsche Forschungsgemeinschaft (DFG, German Research Foundation) through a Major Research Instrumentation grant (project number 492562173) which funded the experimental setup.\\
R. Seiger received funding from the Society for Research Promotion Lower Austria (Gesellschaft für Forschungsförderung Niederösterreich, GFF NÖ) through an Excellence Scholarship for Research (ExzF-0005).

\section*{Author contributions}
P.W., M.H., F.B., J.E., P.Fi., R.H., A.W.-G., L.W. and P.Z. conceived the study and designed the experiments. 
P.W., M.H., T.J., K.K., F.K., and T.L. contributed to the experimental setup and structural installation of the environment. 
P.W., M.H., F.He., T.J., F.K., R.S., J.V., C.W., and L.W. performed the measurements and acquired the data. 
P.W., F.He., and R.S. developed the needed software and analyzed the data, . 
P.Fi. and F.B. managed the engineering and manufacturing design of the industrial shielding components. 
P.Ew., K.N., and A.W.-G. provided clinical context, medical expertise, and coordinated the ethical approval requirements for biomagnetism applications. 
P.Fi. supervised the project and acquired the funding. 
P.W. wrote the original manuscript draft. 
All authors actively participated in reviewing, editing, and providing critical feedback on the manuscript.

\section*{Competing interests}

The authors declare no competing interests.

\end{document}